\documentclass[sigconf,nonacm]{acmart}
\usepackage[nameinlink,capitalise]{cleveref}
\usepackage{xcolor}
\usepackage{fancyvrb} 
\usepackage{tikz}
\usepackage{enumitem}
\usepackage{float}
\usepackage{microtype} 
\usepackage{makecell}

\usetikzlibrary{arrows.meta,positioning,shapes.geometric,shapes.misc,fit,calc}

\newcolumntype{Y}{>{\raggedright\arraybackslash}X}

\begin{document}

\title[Motivations, Characteristics, and Influence Mechanisms of Crypto Key Opinion Leaders]{Credibility Matters: Motivations, Characteristics, and Influence Mechanisms of Crypto Key Opinion Leaders}
\author{Alexander Kropiunig}
\affiliation{\institution{Complexity Science Hub}\city{Vienna}\country{ Austria}}
\email{kropiunig@csh.ac.at}
\author[]{Svetlana Kremer}
\affiliation{\institution{Austrian Institute of Technology}\city{Vienna}\country{ Austria}}
\affiliation{\institution{Complexity Science Hub}\city{Vienna}\country{ Austria}}
\email{abramova@csh.ac.at}
\author[]{Bernhard Haslhofer}
\affiliation{\institution{Complexity Science Hub}\city{Vienna}\country{Austria}}
\email{haslhofer@csh.ac.at}

\hypersetup{
  pdfauthor={Alexander Kropiunig, Svetlana Kremer, Bernhard Haslhofer},
  pdftitle={Credibility Matters: Motivations, Characteristics, and Influence Mechanisms of Crypto Key Opinion Leaders},
  pdfsubject={Preprint},
  pdfcreator={},
  pdfproducer={},
  pdfkeywords={blockchain, credibility, crypto influencer, cryptocurrency, finfluencer, opinion leader, social media}
}

\begin{abstract}

Crypto Key Opinion Leaders (KOLs) shape Web3 narratives and retail investment behaviour. In volatile, high-risk markets, their credibility becomes a key determinant of their influence on followers. Yet prior research has focused on lifestyle influencers or generic financial commentary, leaving crypto KOLs' understandings of motivation, credibility, and responsibility underexplored. Drawing on interviews with 13 KOLs and self-determination theory (SDT), we examine how psychological needs are negotiated alongside monetisation and community expectations. Whereas prior work treats finfluencer credibility as a set of static credentials, our findings reveal it to be a \emph{self-determined, ethically enacted practice}. We identify four community-recognised markers of credibility: self-regulation, bounded epistemic competence, accountability, and reflexive self-correction. This reframes credibility as socio-technical performance, extending SDT into high-risk crypto ecosystems. Methodologically, we employ a hybrid human--LLM thematic analysis. The study surfaces implications for designing credibility signals that prioritise transparency over hype.
\end{abstract}



\maketitle


\section{Introduction}\label{sec:introduction}

Over recent years, social media influencers have emerged as powerful actors shaping consumer preferences and financial decision-making. Global spending on influencer marketing reached \$24~billion in 2024 \cite{pan2025}, reflecting their growing economic significance. Audiences often emulate their online opinion leaders, and influencer-driven content has been shown to significantly affect consumer behaviour~\cite{libai2025influencer}. This influence increasingly extends into financial markets, where digital communities can shape investment decisions. The case of Reddit's \texttt{WallStreetBets} illustrates this phenomenon, as collective attention drove waves of high-risk retail trades \cite{warkulat2024}.

In the blockchain and cryptocurrency sector, a distinct class of influencers, often termed \emph{Key Opinion Leaders (KOLs)} or ``crypto-influencers,'' has emerged. These individuals command large followings and play a central role in shaping sentiment and investment behaviour. Empirical research shows that crypto influencers' communications can drive short-term cryptocurrency price fluctuations and trading volumes, underscoring their market-moving capacity \cite{Merkley2024,Moser2023}. Unlike mainstream lifestyle influencers, crypto KOLs operate in highly volatile, loosely regulated markets characterised by pseudonymous participation, complex tokenomics, and irreversible transactions. They often hold positions in the assets they discuss, promote or build products around, and communicate with communities that are both financially and ideologically invested. In this setting, a single post about a low-cap token may move prices, expose followers to pump-and-dump schemes or rug pulls, and generate substantial losses for retail investors with limited technical or financial expertise. Credibility is therefore not merely an abstract reputational asset but a determinant of real financial risk.

Crypto KOLs differ from generic financial influencers (finfluencers) and traditional financial advisors. Whereas finfluencers frequently comment on diversified portfolios or macro trends, crypto counterparts position themselves as domain specialists in decentralised finance (DeFi) and Web3 protocols, translating highly technical developments into actionable narratives. They may communicate under pseudonyms while simultaneously leaving public traces on-chain (e.\,g., via visible wallet holdings or governance activity), creating new forms of transparency and conflict of interest. Regulatory guidance for such actors is still emerging and fragmented across jurisdictions, leaving many KOLs to self-interpret obligations around disclosure, suitability, and market integrity. These socio-technical conditions make questions of credibility, ethics, and responsibility especially acute.

Academic attention to finfluencers remains still limited. Prior work has examined how influencers build personal brands and motivations in mainstream contexts~\cite{torhonen2019}, how sponsorship disclosure and perceived authenticity affect persuasion~\cite{boerman2017}, and how finfluencers and celebrity figures can move crypto markets in the short run~\cite{Merkley2024,AnteMuskCrypto2023}. Parallel literatures in human-computer interaction (HCI) and usable security document how cryptocurrency users struggle with risks, scams, and security practices~\cite{Krombholz2016,Mai2020,SasKhairuddin2017,Abramova2021}. However, there is still a limited understanding about how crypto KOLs become and remain opinion leaders in such high-stakes environments, how they conceptualise and cultivate credibility, and how they navigate ethical tensions between education, promotion, and self-interest.

Guided by self-determination theory (SDT) \citep{deci2000} and using qualitative research methods, this study examines the motivations, beliefs, and communicative strategies of crypto KOLs. Specifically, we investigate:
\begin{itemize}
  \item \textbf{RQ1:} What extrinsic and intrinsic factors motivate individuals to become and remain key opinion leaders in the cryptocurrency ecosystem?
  \item \textbf{RQ2:} What defining characteristics and everyday practices set crypto KOLs apart from other influencers and members of the crypto community?
  \item \textbf{RQ3:} How do crypto KOLs conceptualise and enact credibility, ethics, and responsibility toward their communities in high-risk, speculative markets?
\end{itemize}

\noindent \textbf{Contributions.} This study advances understanding of crypto-influencer dynamics through four interrelated contributions which are grounded in theory, empirical evidence, methodological reflection, and practical impact.

\emph{Theoretical contribution.} We apply self-determination theory to high-risk financial contexts by showing how the needs for autonomy, competence, and relatedness are negotiated under conditions of volatility, sponsorship pressure, and shared financial exposure with followers. Our analysis surfaces tensions between autonomy and monetisation, between technical competence and communicative accessibility, and between relatedness and the risk of encouraging speculation. We further contribute to influencer marketing and creator-labour research by theorising crypto KOLs as hybrid actors who simultaneously serve as educators, informal financial advisors, entrepreneurs, and community stewards. Whereas prior work conceptualises credibility in terms of static credentials such as formal education, professional background, and financial certifications~\cite{NurhandayaniWhoDeserves2025}, our findings reveal credibility as a \emph{self-determined, ethically enacted practice} rooted in SDT needs and shaped by KOLs' own norms, constraints, and responsibilities. We identify four emergent dimensions that function as community-recognised markers of trustworthiness: (i)~\emph{self-regulation and voluntary constraint}, whereby KOLs decline misaligned sponsorships and impose personal rules on promotion; (ii)~\emph{bounded epistemic competence}, acknowledging the limits of one's expertise and avoiding prognostication; (iii)~\emph{accountability}, cultivating long-term trust through transparent disclosure and community stewardship; and (iv)~\emph{reflexive self-correction}, learning from past failures and continuously reassessing own practices. This reframes credibility as a socio-technical, self-regulated performance rather than a static credential.

\emph{Empirical contribution.} Drawing on thirteen interviews with crypto KOLs from Europe, the United States, and Asia (traders, educators, founders, and analysts), we present a thematic analysis that details (i) extrinsic drivers (e.\,g., sponsorship revenue, monetisation of analytics, social capital), (ii) intrinsic motives (e.\,g., enjoyment, educating newcomers, ideological commitment, desire for mastery and community), (iii) characteristics and practices that distinguish crypto KOLs from mainstream influencers (e.\,g., technical expertise, regulatory literacy, use of on-chain signals), and (iv) ethical and community norms that guide their communication strategies (e.\,g., disclosure practices, self-imposed restrictions on promotions). These qualitative insights fill a notable gap in the literature on digital opinion leadership by providing rich, contextualised accounts of a previously understudied influencer category.

\emph{Methodological contribution.} To systematically analyse a complex and rapidly evolving domain, we describe and critically reflect on a hybrid workflow that couples conventional qualitative interviewing with large language model--assisted thematic analysis. In our approach, a large language model (LLM) proposes candidate codes and SDT mappings on anonymised transcript segments, while human researchers retain full control over the codebook development, theme construction, and interpretation. We show how this human-in-the-loop workflow can broaden candidate theme coverage and support transparency without treating the LLM as an autonomous analyst or source of ground truth.

\emph{Practical contribution.}
Our findings yield actionable recommendations for industry and policy. For marketers, campaigns that resonate with KOLs' community-oriented motives such as education and protection are more likely to produce credible and enduring partnerships than those based solely on one-off paid promotions. For platform designers, we suggest features such as transparent disclosure badges, longitudinal accuracy or ``track record'' indicators, and cross-platform reputation profiles that help users identify credible actors and incentivise responsible behaviour. For regulators, the study highlights the need for proportionate disclosure rules, registration thresholds for high-reach influencers, and cross-jurisdictional coordination to address scams, rug pulls, and other forms of market manipulation while preserving the participatory ethos of Web3. Together, these measures offer a roadmap for strengthening trustworthy influence in cryptocurrency and for designing socio-technical infrastructures that support ethical, evidence-based content.

\noindent \textbf{Paper structure.} The remainder of this paper is organised as follows. Section~\ref{sec:background} reviews relevant literature on cryptocurrency user behaviour, social media influencers, finfluencers, creator labour, and the SDT framework. Section~\ref{sec:methodology} describes our qualitative methodology, including interview procedures, participant demographics, and coding strategy. Section~\ref{sec:results} presents the main results structured along the specified RQs. In Section~\ref{sec:discussion}, we discuss the implications of these findings, limitations and future research directions, and Section~\ref{sec:conclusion} concludes with a summary of contributions.


\section{Background and Related Work}\label{sec:background}

This section synthesises prior research to situate crypto key opinion leaders (KOLs) as a distinct category of digital workers. We organise the literature around four themes: (1) influencers, finfluencers, and creator labour; (2) why crypto constitutes a qualitatively different context; (3) crypto KOLs as digital creators under financialised conditions; and (4) the research gap motivating our study.

\subsection{Influencers, Finfluencers, and Creator Labour}

\subsubsection{Influencer Culture and Creator Labour}

Influencer research conceptualises social media creators as micro-celebrities who strategically curate authenticity and intimacy to attract audiences \cite{MarwickStatusUpdate2013,Khamis2017,Enke2019}. Ethnographic accounts reveal how influencers cultivate personas that are simultaneously relatable and aspirational, with authenticity understood as performative, carefully crafted to sustain attention and trust \cite{MarwickStatusUpdate2013,BaymPlaying2018}.

Two concepts capture the digital labour involved. Abidin's \emph{visibility labour} describes the ongoing work of posting, curation, and interaction through which influencers maintain presence \cite{AbidinVisibility2016}. Baym's \emph{relational labour} emphasises the effort invested in sustaining intimate audience relationships \cite{BaymPlaying2018}. Both are consequential because parasocial interaction makes followers receptive to recommendations; credibility emerges from affective bonds, not just expertise \cite{Lou2019,Labrecque2014,Reinikainen2020}.

From a political-economy perspective, Duffy and colleagues describe  ``nested precarities'' of creators, which denotes overlapping insecurities at the level of platforms, industries, and algorithms \cite{DuffyPlatformPractices2019,DuffyNestedPrecarities2021}. Creators engage in constant self-monitoring, cross-posting, and brand management to remain visible \cite{DuffyAspirationalWork2017}, ``playing the visibility game'' to satisfy opaque ranking systems \cite{CotterVisibilityGame2019}. Success depends on algorithmic regimes and revenue streams that can shift abruptly.

\subsubsection{Finfluencers and Social-Media Finance}

Recent scholarship extends influencer culture into personal finance. Guan characterises ``finfluencers'' as online financial advice-givers who mix analysis, storytelling, and entertainment, yet often operate outside regulatory frameworks \cite{GuanFinfluencer2023}. Hayes and Ben-Shmuel show how finfluencers contribute to the financialisation of everyday life by weaving investing into lifestyle narratives, normalising speculative behaviours \cite{HayesFinfluence2024}. This work highlights diverse motivations, ranging from education and community building to self-promotion, and identifies conflicts of interest when creators profit from endorsed products.

Quantitative studies demonstrate that finfluencer activity shapes crowd sentiment. Haase et al.\ find that finfluencers' sentiment Granger-causes broader crowd sentiment, particularly in crypto discussions \cite{HaaseFinfluencersSentiment2025}. Meyer et al.\ document emotional contagion in YouTube crypto content, with influencers' tone mirrored in audience comments \cite{MeyerHighOnBitcoin2023}. These findings suggest finfluencers modulate affect, amplifying herd dynamics in volatile markets.

\subsubsection{Credibility in Finfluencer Research}

Existing research on finfluencer credibility has predominantly adopted a \emph{credential-based} perspective, conceptualising trustworthiness in terms of static attributes such as formal education, professional background, financial certifications, and practitioner experience~\cite{NurhandayaniWhoDeserves2025}. This framing treats credibility as a property that individuals possess by virtue of their qualifications, implicitly privileging institutional markers over situated practices. However, such an approach may not fully capture how credibility operates in decentralised, pseudonymous environments like crypto, where formal credentials are often absent, unverifiable, or irrelevant to community norms. In these contexts, credibility may instead emerge from observable behaviours (such as transparent disclosure, restraint in promotion, and responsiveness to community feedback) rather than from static signals of expertise. This gap motivates our inquiry into how crypto KOLs \emph{enact} credibility through self-determined practices, an approach we analyze empirically in Section~\ref{sec:rq3} and theorise in Section~\ref{sec:discussion}.

\subsubsection{Creator Labour Under Financialisation}

Duffy's work on \emph{aspirational labour} shows how creators invest substantial unpaid time building visibility in hope of future opportunities \cite{DuffyAspirationalWork2017}. Financialisation intensifies these dynamics. Alacovska and Chalcraft analyse NFT artists' work as ``speculative labour,'' where creative outputs become financial assets subject to extreme volatility \cite{AlacovskaSpeculativeLabour2024}. Internet celebrity dynamics \cite{AbidinInternetCelebrity2018} illustrate how rapid visibility generates both opportunities and vulnerability to backlash when advice fails. In such environments, creators' incomes, reputations, and portfolios become entangled with promoted products.

This literature establishes that influencers are not merely communicators but \emph{workers} under conditions of nested precarities. Crypto KOLs inherit these dynamics but operate in a high-risk financial domain where attention fluctuations translate directly into gains or losses, foregrounding questions of credibility and responsibility.

\subsection{Crypto as a Distinct Context}

Cryptocurrency presents a qualitatively different context for influencer work. We synthesise research on usability, regulation, scams, and community trust to explain why crypto KOLs occupy a particularly consequential position.

\subsubsection{Fragile Mental Models and Usability Struggles}

HCI research underscores that cryptocurrencies remain difficult to use safely. Many users struggle with private-key management and rely on third-party services, exposing them to custodial risk \cite{MoserBitcoinUsability2013,Krombholz2016}. Studies reveal fragile mental models of custody, anonymity, and transaction finality, including assumptions about reversibility that rarely hold \cite{Mai2020,Abramova2021}. This suggests that KOL audiences often lack foundational knowledge to critically evaluate advice.

\subsubsection{Regulatory Gaps and Weak Consumer Protections}

Cryptocurrency promotion falls between or outside conventional investor-protection regimes. Finfluencers' undisclosed sponsorships and pump-and-dump schemes evade rules designed for licensed advisors \cite{StefanouFinfluencersReg2022}. Regulatory agencies have issued guidelines targeting misleading promotions \cite{SECInvestorCrypto,FCA2023Finfluencer,ASAInfluencer2023}, yet enforcement remains patchy and jurisdictionally fragmented. Crypto KOLs often operate in even murkier environments, promoting speculative assets that cross national boundaries and challenge securities law. This intensifies both KOLs' impact and ethical tensions around disclosure.

\subsubsection{Scams and Community Trust}

Oak and Shafiq analyse scam-related discourse on Reddit, identifying roles of victims, vigilantes, and advice-givers that underscore grassroots demand for trustworthy guidance \cite{OakScamReddit2025}. On-chain analyses reveal ecosystems of scam tokens and rug pulls that exploit naive investors \cite{Cernera2023rugpulls}. Research on user trust highlights a ``centralised trust in decentralised systems'' paradox: users anchor trust in exchanges and charismatic community figures, precisely the intermediaries decentralisation was meant to obviate \cite{KhairuddinBitcoinMotivations2016,SasKhairuddin2017,BappyCentralizedTrust2025}.

Together, these strands establish that crypto is a socio-technical environment where users struggle with operations, protections are weak, and scams are pervasive. Users often ``outsource'' due diligence to community figures, including KOLs whose endorsements function as trust shortcuts, akin to, but less regulated than, traditional financial advisors.

\subsection{Crypto KOLs as Digital Workers}

We now reframe crypto KOLs as a form of digital creator labour at the intersection of influencer culture and crypto's specific risks.

\subsubsection{Double Exposure: Platform and Market Precarity}

Crypto KOLs exemplify intersecting forms of precarity. Like other creators, they depend on platform algorithms and constant content production \cite{DuffyPlatformPractices2019,CotterVisibilityGame2019}, engaging in visibility and relational labour \cite{AbidinVisibility2016,BaymPlaying2018} while navigating nested precarities \cite{DuffyNestedPrecarities2021}.

However, unlike lifestyle influencers, crypto KOLs' success is tied directly to volatile asset prices. Their work constitutes speculative labour \cite{AlacovskaSpeculativeLabour2024}, with reputations and portfolios entangled with promoted products. Lacking institutional safeguards or licensing, they face a \emph{double exposure} to both platform dynamics and financial markets, rendering their labour uniquely precarious. Internet celebrity dynamics \cite{AbidinInternetCelebrity2018} apply forcefully: rapid visibility generates opportunities but also vulnerability to backlash when projects fail.

\subsubsection{Self-Determination Theory and KOL Motivations}

SDT provides a framework for understanding motivations beyond ``greed'' versus ``altruism.'' Central to SDT is a continuum of relative autonomy, ranging from amotivation through external regulation to fully internalised, intrinsic motivation; the degree to which external incentives become self-endorsed determines the quality and persistence of engagement \cite{RyanDeci2000}. SDT also posits three basic psychological needs: \emph{autonomy}, \emph{competence}, and \emph{relatedness}, whose satisfaction supports this internalisation process and fosters well-being \cite{DeciRyan1985,RyanDeci2000}.

SDT has been applied to online communities and digital work. Nov finds Wikipedians driven by intrinsic enjoyment, learning, and prosocial motives \cite{NovMotivatesWikipedians2007}. Tyack and Mekler show that supporting SDT needs fosters sustained engagement \cite{TyackMeklerSDT2020}. Studies of content creators find that YouTube producers value autonomy, competence, and relatedness, with monetary rewards often secondary \cite{LiuAnxietyCommunities2021,torhonen2019}.

Applying SDT to crypto KOLs, we conceptualise how they may experience \emph{autonomy} in choosing projects, \emph{competence} in decoding protocols for lay audiences, and \emph{relatedness} in cultivating communities. Simultaneously, sponsorships and portfolio performance exert external pressures. An SDT lens distinguishes internalised motivations (education, community stewardship) from controlled ones (sponsor obligations, trading gains), providing a framework for analysing credibility practices.

\subsubsection{Market-Level Influence}

Research links online influence to financial outcomes. Studies of r/WallStreetBets show how social dynamics moved markets during the GameStop squeeze \cite{LucchiniGamestopWSB2022,warkulat2024}. In crypto, Musk's tweets produce significant price changes \cite{AnteMuskCrypto2023}, and finfluencer endorsements channel capital into tokens \cite{Merkley2024,HayesFinfluence2024,HaaseFinfluencersSentiment2025,MeyerHighOnBitcoin2023}.

Across this literature, influencers are modelled as sources of signals correlated with market movements. Much less attention is paid to their experiences as workers, or how motivations and ethical commitments shape their practices.

\subsubsection{Conceptual Summary: Creator-Labour and SDT Concepts Applied to Crypto KOLs}

Table~\ref{tab:concepts-kols} summarises how core concepts from creator-labour scholarship and SDT map onto the work of crypto KOLs. This conceptual mapping guides our empirical analysis by identifying the forms of labour, precarity, and motivation that characterise KOL work.

\begin{table*}[htbp]
\centering
\caption{Conceptual framework mapping creator-labour constructs and SDT needs to the work of crypto key opinion leaders.}
\label{tab:concepts-kols}
\small
\renewcommand{\arraystretch}{1.2}
\begin{tabular}{@{} >{\raggedright\arraybackslash}p{3.5cm} >{\raggedright\arraybackslash}p{6.2cm} >{\raggedright\arraybackslash}p{6.2cm} @{}}
\toprule
\textbf{Concept} & \textbf{Definition} & \textbf{Application to Crypto KOLs} \\
\midrule
\textit{Creator-labour constructs} & & \\[-4pt]
\cmidrule(r){1-1}
\textbf{Visibility labour} {\cite{AbidinVisibility2016}} & Ongoing work of posting, curation, and interaction to maintain public presence & Continuous content production, audience engagement, and cross-platform presence management \\[6pt]
\textbf{Relational labour} {\cite{BaymPlaying2018}} & Effort invested in building and maintaining intimate audience relationships & Cultivating parasocial bonds through community engagement and sharing personal experiences \\[6pt]
\textbf{Aspirational labour} {\cite{DuffyAspirationalWork2017}} & Unpaid or under-compensated work invested in hope of future opportunities & Heavy early-career investment in content and community before monetisation \\[6pt]
\textbf{Nested precarities} {\cite{DuffyNestedPrecarities2021}} & Overlapping insecurities at platform, industry, and economic levels & Algorithmic volatility, market crashes, sponsorship instability, and regulatory uncertainty \\[6pt]
\textbf{Speculative labour} {\cite{AlacovskaSpeculativeLabour2024}} & Creative work entangled with volatile financial assets & Reputations and incomes tied to token prices; portfolios often include promoted projects \\[6pt]
\textbf{Internet celebrity} {\cite{AbidinInternetCelebrity2018}} & Rapid visibility generating opportunities and vulnerabilities & Successful calls attract followers and sponsorships; failed calls invite backlash \\[3pt]
\addlinespace[4pt]
\textit{SDT basic needs}$^{\dagger}$ & & \\[-4pt]
\cmidrule(r){1-1}
\textbf{Autonomy} & Need to feel self-directed and volitional & Freedom to choose projects, content style, and community norms \\[6pt]
\textbf{Competence} & Need to feel effective and capable & Satisfaction from accurate analyses, successful predictions, and audience learning \\[6pt]
\textbf{Relatedness} & Need for meaningful social connections & Community ties and peer networks; prosocial motives coexist with commercial ones \\
\bottomrule
\addlinespace[2pt]
\multicolumn{3}{@{}l}{\footnotesize$^{\dagger}$\,All three needs drawn from the self-determination theory~\cite{DeciRyan1985,RyanDeci2000}.}
\end{tabular}
\end{table*}

\subsection{Research Gap}

A comprehensive mapping of the literatures, their relevance to crypto KOLs, and remaining gaps is provided in \Cref{tab:lit-synthesis} (Appendix). Despite this rich body of work, a crucial gap remains. Existing research focuses on \emph{crypto users and retail investors}, examining their struggles and trust practices; on \emph{market-level consequences}, modelling influencers as sentiment signals; or on \emph{lifestyle influencers and generic finfluencers}, without attending to crypto's specific conditions. We lack an in-depth, SDT-informed qualitative account of \emph{crypto KOLs themselves}.

This gap matters because crypto KOLs occupy a critical, under-researched position as community figures whose motivations and practices have material consequences for followers' financial well-being. Understanding KOLs as workers, rather than merely as market signals, is essential for theory, intervention design, and policy. Our research questions (Section~\ref{sec:introduction}) address this gap: KOL motivations (RQ1), distinguishing practices (RQ2), and credibility and ethical reasoning (RQ3).

\section{Methodology}\label{sec:methodology}

We employed a qualitative design comprising semi-structured interviews with thirteen crypto KOLs and a hybrid thematic analysis guided by self-determination theory (SDT). Two trained researchers conducted and coded the interviews, and we used a large language model in a human-in-the-loop workflow to propose additional candidate codes that were subsequently curated and verified by humans. We assessed inter-annotator reliability between the two human coders and integrated the resulting themes into an SDT-aligned framework. Detailed procedures follow.
\subsection{Ethical Considerations}
\label{subsec:ethics}
The study protocol was reviewed and approved by the TU Wien Research Ethics Committee (TUW REC, Case No.\ 080/27062025). All participants were adults ($\geq$18 years) and provided informed consent. Safeguards included strict pseudonymization (e.\,g., ``KOL01''), removal of direct identifiers in transcripts and publications, and GDPR-compliant retention limits (audio deleted after one year; anonymized transcripts after three). Participants could skip questions, pause, or withdraw until October 31, 2025. Quotations were anonymized and cleaned prior to publication, research materials were stored securely, and protocols governed the handling of sensitive disclosures to ensure confidentiality unless imminent harm was reported. For language editing, large language models were used only with fully anonymized excerpts containing no identifiers; no raw data were transferred externally, and all core processing occurred within the controlled research environment.

\paragraph{Model use and additional safeguards.}
We used OpenAI GPT-4 as a supplementary assistant for candidate theme suggestion (temperature~=~0.2; top\_p~=~1.0). Before any model input, all processed excerpts were fully anonymized. Only short, de-identified segments were processed; no raw datasets or identifiers left the research environment. All model outputs were reviewed by human coders and accepted, revised, or discarded.

\subsection{Qualitative Interviews}

We conducted 13 semi-structured interviews with Key Opinion Leaders (KOLs) within the blockchain ecosystem. Participants were recruited through targeted sampling of individuals who met specific criteria: (1) active presence on social media platforms with crypto-focused content, (2) demonstrated influence within the crypto community through follower engagement or industry recognition, and (3) consistent content creation related to blockchain technology, cryptocurrency, or decentralized finance. Initial participants were identified through systematic searches of prominent social media accounts, and additional participants were recruited through snowball sampling based on referrals from initial interviewees.

Each interview lasted between 30 and 60 minutes and was conducted via video conferencing platforms to accommodate participants' global distribution. Interview questions were organised around three main areas: (1) motivations for becoming a crypto KOL, (2) strategies for building and maintaining influence, and (3) perspectives on ethics and responsibility within the crypto community. All interviews were recorded with participant consent and transcribed verbatim for thematic analysis.

Building on this design, we implemented an interview protocol explicitly aligned with our research questions (RQs) to elicit rich narratives while preserving comparability across the interviews. The interview questions are provided in \Cref{app:interview-questions}. Each question also includes an identifier mapping it to the associated RQ(s).


\subsection{Participant Profile}
\label{subsec:participant-profile}

The study examined a heterogeneous cohort of key opinion leaders (KOLs), ranging from early crypto adopters to post-2020 entrants shaped by DeFi and Web3 infrastructures. Participants occupy hybrid roles: educator, builder, researcher, and entrepreneur, combining academic training with market practice. This hybridity positions KOLs as intermediary nodes that translate between technical discourse, retail education, and regulatory concerns. Geographically, participants were based in Europe, the United States, and Asia.

Platform repertoires are multi-sited. Core channels include X (formerly Twitter), LinkedIn, YouTube, and Telegram, often complemented by Instagram, TikTok, and specialized academic or DAO channels. Content formats span short-form explainers, livestreams, research threads, technical updates, and policy commentary. 

Table~\ref{tab:kols} synthesises entry timing, roles, content orientation, platform portfolios, professional backgrounds, and topical foci. The distribution reveals three overlapping archetypes: (i) \emph{educator-curators} standardising retail-facing knowledge; (ii) \emph{researcher-practitioners} linking academic and technical insights to market narratives; and (iii) \emph{builder-entrepreneurs} embedding communication within product and ecosystem development. These archetypes recur across platforms, indicating that influence is less channel-specific than orchestrated across multiple venues.

\begin{table*}[t] 
\centering
\caption{Anonymized overview of KOLs: entry timing, self-defined role, content style, platforms, background, and focus. \newline Abbrev.: YT=YouTube, TG=Telegram, X=X (formerly Twitter), IG=Instagram, TT=TikTok, LI=LinkedIn.}
\label{tab:kols}
\footnotesize
\setlength{\tabcolsep}{3pt} 
\renewcommand{\arraystretch}{1.2} 

\begin{tabular}{@{} l l p{2.9cm} p{2.9cm} c c c c c c p{2.9cm} p{2.9cm} @{}}
\toprule
\multicolumn{4}{c}{\textsc{\footnotesize Profile}} &
\multicolumn{6}{c}{\textsc{\footnotesize Platform}} &
\multicolumn{2}{c}{\textsc{\footnotesize Context}} \\
\cmidrule(r){1-4} \cmidrule(lr){5-10} \cmidrule(l){11-12}
\textbf{\small KOL} & \textbf{\small Entry} & \textbf{\small Role} & \textbf{\small Content} &
\textbf{\small YT} & \textbf{\small TG} & \textbf{\small X} & \textbf{\small IG} & \textbf{\small TT} & \textbf{\small LI} &
\textbf{\small Background} & \textbf{\small Focus} \\
\midrule
KOL-01 & 2019 & Educator (finance) & Videos, tips & \checkmark & \checkmark & \checkmark & \checkmark & \checkmark &  & Coaching / studies & Retail investing / education \\
KOL-02 & 2015 & Educator & Short videos, news & \checkmark &  & \checkmark & \checkmark & \checkmark &  & Insurance & Crypto education \\
KOL-03 & 2015 & Advocate / educator & Essays, commentary &  &  &  &  &  & \checkmark & Research / arts & Blockchain advocacy \\
KOL-04 & 2017 & Researcher / educator & Research, lectures &  &  &  &  &  & \checkmark & Academic background & Crypto-economic research \\
KOL-05 & 2017 & Educator / researcher & Articles, scholarly & \checkmark &  &  &  &  & \checkmark & Academic + industry & Teaching / research \\
KOL-06 & 2016 & Builder / founder & Technical updates &  & \checkmark & \checkmark &  &  &  & Tech background & DeFi infrastructure \\
KOL-07 & pre-2016 & Educator / influencer & Daily posts, newsletter & \checkmark &  & \checkmark &  &  & \checkmark & Finance / teaching & Retail finance education \\
KOL-08 & 2016 & Builder / academic & Research, commentary &  &  & \checkmark &  &  & \checkmark & Markets / tech & DeFi \& DAO research \\
KOL-09 & 2017 & Educator / influencer & Tutorials, guides & \checkmark & \checkmark & \checkmark & \checkmark &  &  & Entrepreneurial & Web3 education \\
KOL-10 & 2015 & Builder / founder & Startup commentary &  &  & \checkmark &  &  &  & Tech / consulting & Web3 products \\
KOL-11 & 2015 & Entrepreneur / expert & Webinars, blogs &  &  &  &  &  &  & Tech / auditing & Regulation / compliance \\
KOL-12 & 2021 & Investor / mentor & Investment insights &  &  &  &  &  & \checkmark & Entrepreneurship / VC & Web3 investment \\
KOL-13 & 2018 & Researcher / investor & Research threads &  &  & \checkmark &  &  & \checkmark & Tech / finance & Crypto ventures \\
\bottomrule
\end{tabular}

\end{table*}

All participants were adults. To reduce re-identification risks in this relatively small and specialised population, we do not report more granular demographic breakdowns (e.\,g., precise age ranges or gender identities) beyond the regional and role information summarised in Table~\ref{tab:kols}. Participants first engaged with cryptocurrencies between 2015 and 2021 (Table~\ref{tab:kols}), meaning that at the time of interview they spanned early adopters with nearly a decade of experience and more recent entrants who became active during the expansion of DeFi and Web3 infrastructures.

To balance both depth and diversity while remaining feasible within our resources, we aimed for a sample of approximately a dozen KOLs. Recruitment and analysis proceeded iteratively; after the thirteenth interview, new accounts primarily reinforced existing patterns in motivations, practices, and ethical reflections, and we therefore judged that we had reached theoretical saturation for our focal population. This sampling strategy, however, introduces several limitations. Participants were drawn from Europe, the United States, and Asia, and produced content on major global platforms (e.\,g., X, YouTube, LinkedIn, Telegram), which under-represents KOLs from other regions, language communities, or platforms. Moreover, most interviewees already had substantial audiences and multi-platform presences, so our findings may overstate the perspectives of relatively established and successful KOLs compared to smaller or emerging creators. These biases constrain the generalisability of our findings to the broader population of crypto content creators; we return to these limitations in Section~\ref{sec:discussion}.

\subsection{LLM-Assisted Thematic Analysis}

\paragraph{Rationale for LLM assistance.}
We introduced LLM support to explore on a small interview corpus whether model-suggested candidate themes could broaden the space of codes considered in standard qualitative analysis without reducing interpretive quality. The aims were to (i) broaden multi-label coverage, (ii) improve consistency and auditability of the working codebook, and (iii) accelerate SDT-oriented theory linkage. In line with our ethics protocol, the LLM was strictly supplementary: all suggestions were treated as hypotheses, reviewed by human coders against the raw text, and retained only when clearly grounded in participant accounts.

Our analysis employed a hybrid thematic analysis that combined traditional human-led analysis with LLM-suggested candidate themes to systematically identify and refine patterns across the interview transcripts. This approach leveraged the analytical expertise of trained qualitative researchers while using the pattern-recognition capabilities of large language models to surface additional codes for human consideration. We used OpenAI GPT-4 (June 2025) with low-temperature settings (temperature~=~0.2, top\_p~=~1.0); prompts and example outputs are documented in Appendix~\ref{app1}, Section~2 (\Cref{lst:ai-prompts,lst:ai-output}).

\begin{figure*}[!t]
  \centering
\begin{tikzpicture}[x=.05cm, node distance=4mm and 12mm,
 >=Stealth,
  every node/.style={font=\footnotesize},
  process/.style={
    rectangle, rounded corners=2pt, draw,
    line width=0.4pt, align=center,
    minimum height=9mm, text width=20mm, inner sep=2.4mm,
    fill=gray!3
  },
  data/.style={process, fill=gray!12},
  groupbox/.style={draw, rounded corners=2pt, inner sep=5mm, dashed},
  forward/.style={->, line width=0.6pt},
  assist/.style={->, densely dotted, line width=0.6pt},
  feedback/.style={->, dashed, line width=0.5pt}]
  \node[data]    (seg)      {Transcript\\Segmentation};
  \node[process, right=4mm of seg]      (human)    {Dual Human\\Descriptive Coding};
  \node[process, above=of human]    (llm)      {LLM\\Candidate Themes};
  \node[process, right=4mm of human] (integrate) {Theme Integration\\\& Reconciliation\vphantom{Aq}};
  \node[process, right=4mm of integrate, text width=22mm] (refine)   {Iterative Refinement\\\& Validation};
  \node[process, right=4mm of refine]   (apply)    {Apply Refined\\Thematic Frame};
  \node[data,    right=4mm of apply]    (themes)   {Themes \& SDT\\Mapping};

  \draw[forward] (seg) -- (human);
  \draw[forward] (human) -- (integrate);
  \draw[assist]  (llm) -- (integrate);
  \draw[forward] (integrate) -- (refine);
  \draw[forward] (refine) -- (apply);
  \draw[forward] (apply) -- (themes);

  \draw[feedback, bend right=20] (integrate.north) to (llm.east);
  \draw[feedback, bend left=15] (integrate.south) to (human.south);
  \draw[feedback, bend left=20] (apply.south)     to (human.south);

  \node[groupbox, fit=(seg)(human), inner sep=4pt, label={[align=center]above left:Human Coding}] (gHuman) {};
  \node[groupbox, fit=(llm), inner sep=4pt, label={[align=center]above:LLM Assistance}] (gLLM) {};
  \node[groupbox, fit=(integrate)(refine)(apply), inner sep=4pt, label={[align=center]above:Synthesis \& Validation}] (gSyn) {};
  \node[groupbox, fit=(themes), inner sep=4pt, label={[align=center]above:Outcomes}] (gOut) {};

  \coordinate (leg) at ($(seg.south west)+(0,-11mm)$);
  \node[anchor=west] at (leg) {\textbf{Legend}};
  \draw[forward] ($(leg)+(0,-3.5mm)$) -- ++(8mm,0);
  \node[anchor=west] at ($(leg)+(9mm,-3.5mm)$) {Primary flow};
  \draw[assist]  ($(leg)+(0,-7.0mm)$) -- ++(8mm,0);
  \node[anchor=west] at ($(leg)+(9mm,-7.0mm)$) {LLM-assisted input};
  \draw[feedback] ($(leg)+(0,-10.5mm)$) -- ++(8mm,0);
  \node[anchor=west] at ($(leg)+(9mm,-10.5mm)$) {Feedback / iteration};
\end{tikzpicture}
  \caption{LLM-assisted thematic analysis workflow combining human analysis and LLM-suggested candidate themes.}
  \Description{A flowchart showing the iterative coding workflow with three main phases: initial human coding of interview transcripts, LLM-assisted code generation and refinement, and final verification and consolidation by human researchers. The diagram shows feedback loops between phases to ensure comprehensive thematic coverage.}
  \label{fig:ai-coding-flow}
\end{figure*}
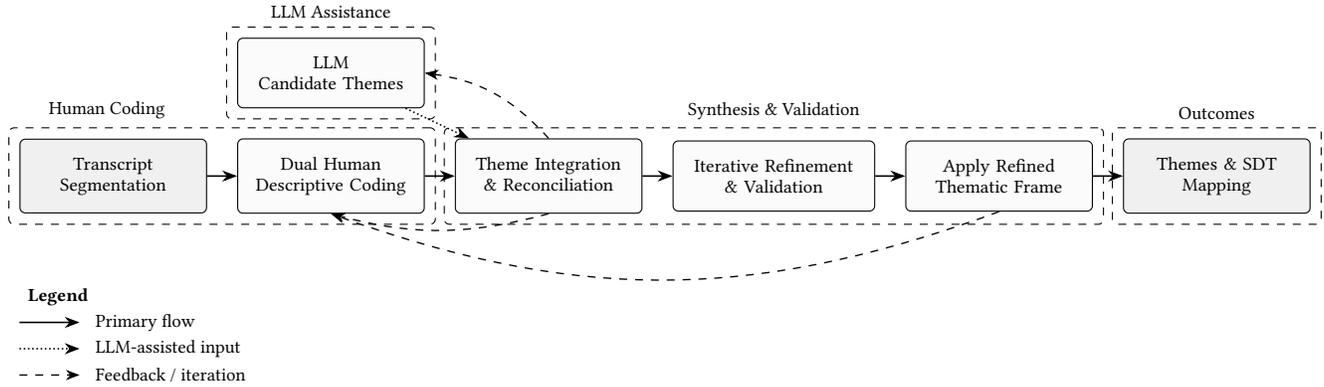

The coding process followed a multi-stage approach (as illustrated in Fig.~\ref{fig:ai-coding-flow}):

\textbf{Stage 1: Transcript Segmentation.} Interview transcripts were divided into segments of ca. 200--300 words. Segment boundaries were aligned with natural conversational pauses and thematic shifts to preserve coherence. This approach provided sufficient detail for fine-grained analysis while retaining the broader context of participants' narratives.

\textbf{Stage 2: Dual Human Descriptive Coding.} Two trained qualitative researchers independently annotated all transcript segments using descriptive coding techniques. Both analysts had experience in qualitative methodology and were familiar with self-determination theory and influencer marketing literature. They worked independently to identify key concepts, motivations, behaviours, and attitudes expressed within each segment, generating candidate thematic labels that remained close to participants' language.

\textbf{Stage 3: LLM-Suggested Candidate Themes.} In parallel to human analysis, each transcript chunk was processed with a large language model (OpenAI GPT-4) using prompts designed to suggest additional candidate themes and potential SDT mappings that might capture patterns not immediately apparent to human analysts. In the first pass, the LLM suggested approximately 60 \emph{unique} labels. The research team then inspected this list, removed obvious duplicates, collapsed near-synonyms, and discarded labels that moralised participants' accounts, introduced speculative psychological diagnoses, or were insufficiently grounded in the underlying text. This process yielded 32 candidate labels for further consideration. The exemplary prompt as well as an excerpt of the model's JSON output structure are provided in Appendix~\ref{app1}, Section~2 (see \Cref{lst:ai-prompts,lst:ai-output}).

\textbf{Stage 4: Theme Integration and Reconciliation.}
The research team systematically compared human-generated labels with the filtered set of LLM-suggested candidates and incorporated those that added analytical depth or coverage. This process resulted in a combined set of 51 labels in total. Of these, 32 originated from LLM suggestions that survived human curation, while 19 were generated independently by the two human coders. Discrepancies between human analysts were resolved through discussion rounds and consensus, with LLM suggestions serving solely as supplementary inputs during reconciliation. This process yielded a comprehensive pool that blended human interpretive expertise with machine-assisted pattern recognition while keeping humans in full control of coding decisions.

\textbf{Stage 5: Iterative Refinement and Validation.}
Through iterative reduction, the pool of 51 labels was distilled into five \emph{overarching themes} and sixteen \emph{subthemes}. The integrated set underwent multiple rounds of refinement: (1) labels were analysed for patterns and redundancies, (2) similar labels were consolidated for conceptual clarity and distinctiveness, (3) the refined thematic framework was reapplied to transcript segments to test consistency and coverage, and (4) final validation was achieved through manual review of supporting quotations to confirm accuracy. Throughout this process, only human-generated and human-curated labels entered the final framework; LLM outputs functioned exclusively as inputs to early coding discussions. This procedure produced a stable and transparent thematic framework, grounded in both systematic procedure and qualitative standards.

\paragraph{Bias and hallucination mitigation.}
At each stage, we treated LLM outputs as tentative hypotheses rather than authoritative analyses. Candidate labels that could not be substantiated by multiple excerpts, that relied on speculative attributions of intent, or that introduced moralising language were removed. We also monitored for systematic differences in how the model described participants from different regions or roles, revising or discarding labels that appeared to reflect such biases. For instance, we rejected codes such as ``LinkedIn for Professional Use'' (too platform-specific to generalise), ``Critical Evaluation of Influencers'' (conflated self-reflection with external critique), and ``Industry Credibility and Expertise'' (attributed domain-level authority without textual grounding). Each rejection was logged with a brief rationale to maintain an auditable curation trail; the complete code curation pipeline, including rejection categories and counts, is documented in Appendix~\ref{app:code-curation}. Final themes and interpretations were based solely on human review of the transcripts and the curated codebook.
%





\subsection{Inter-Annotator Reliability}

We assessed inter-annotator reliability for the two human coders on a stratified random sample of 120 transcript segments drawn across all interviews. Annotation was \emph{multi-label}: each segment could receive zero, one, or multiple candidate themes from the working label set. Reliability was computed as label\,$\times$\,segment binary decisions (presence/absence) per annotator, and we report agreement using Krippendorff's alpha (nominal metric), which is appropriate for multi-label categorical annotation.

Pairwise agreement metrics for the two human coders are reported in \Cref{tab:agreement}. Across the working set of 36 candidate themes and 120 segments, Krippendorff's $\alpha = 0.78$, indicating substantial reliability for the final codebook. The LLM was not treated as an annotator; its suggestions were used only upstream to expand the candidate label set before human coding.

\begin{table}[H]
\centering
\caption{Pairwise inter-annotator agreement under multi-label annotation across 36 candidate themes for the two human coders. Agreement percentage and Krippendorff's alpha (nominal) are reported for the annotator pair.}
\label{tab:agreement}
\begin{tabular}{lcc}
\toprule
\textbf{Pairwise Comparison} & \textbf{\% Agreement} & \textbf{Krippendorff's $\alpha$} \\
\midrule
Anno. 1 vs Anno. 2 & 82\% & 0.78 \\
Anno. 1 vs LLM-assisted & 58\% & 0.55 \\
Anno. 2 vs LLM-assisted & 62\% & 0.59 \\
\bottomrule
\end{tabular}
\end{table}

\subsection{Theme Construction and SDT Integration}

Following the coding stages described above, the human research team conducted thematic analysis guided by self-determination theory (SDT) to organise labels into coherent themes addressing our research questions. The analysis proceeded through three phases: (1) \textbf{descriptive review} to identify surface-level patterns and explicit motivations, (2) \textbf{interpretive synthesis} to uncover deeper psychological needs and implicit drivers, and (3) \textbf{theoretical integration} to connect emergent themes with SDT's core constructs of motivation, autonomy, competence, and relatedness. Throughout these phases, we engaged in iterative memo writing and team discussions to challenge early interpretations and to clarify how themes mapped onto RQ1--RQ3 and the SDT needs.

For an overview of the synthesis, see Appendix~\ref{app1} (Table~\ref{tab:thematic-tree}), which links overarching themes and subthemes to the research questions and provides concise definitions and representative quotations.


\section{Results}\label{sec:results}

Our thematic analysis of 13 interviews with crypto KOLs produces three sets of empirical findings aligned with the research questions. We interpret the recurring patterns and characteristics through the lens of SDT, providing conceptual grounding. Themes and subthemes are summarized in \Cref{tab:thematic-tree} with representative quotes and RQ mapping.

\subsection{RQ1: Motivational Factors for Crypto KOLs}\label{sec:rq1}

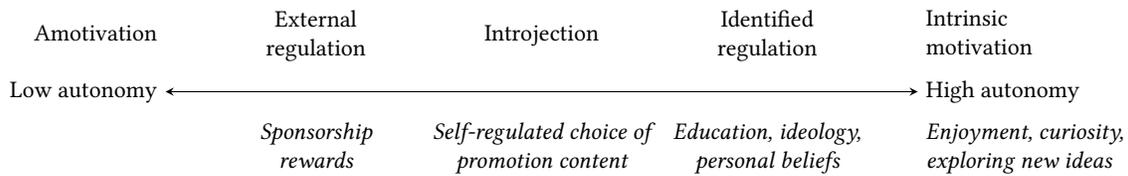
\begin{figure*}[ht]
    \centering
    \begin{tikzpicture}[>=stealth]
        \draw[<->] (0,0.25) node[left]{Low autonomy} -- (10,0.25) node[right]{High autonomy};
        \node[left] at (0,1) {Amotivation\vphantom{Aq}};
        \node[align=center] at (2,1) {External\\regulation};
        \node[align=center] at (5,1) {Introjection};
        \node[align=center] at (8,1) {Identified\\regulation};
        \node[align=left, right] at (10,1) {Intrinsic\\motivation\vphantom{Aq}};
        \node[align=center,font =\itshape] at (2,-.5) {Sponsorship\\rewards\vphantom{Aq}};
        \node[align=center,font =\itshape] at (5,-.5) {Self-regulated choice of\\ promotion content\vphantom{Aq}};
        \node[align=center,font =\itshape] at (8,-.5) {Education, ideology,\\personal beliefs\vphantom{Aq}};
        \node[align=left, right,font =\itshape] at (10,-.5) {Enjoyment, curiosity, \\exploring new ideas\vphantom{Aq}};
    \end{tikzpicture}
    \caption{Motivational factors for crypto KOLs (\emph{in italics}) through the lens of SDT.}
    \label{fig:motivations}
    \Description{Motivational factors for crypto KOLs are mapped to SDT constructs and positioned along a continuum of relative autonomy (from low to high): sponsorship rewards (external regulation); self-regulated choice of promotion content (introjection); education, ideology, personal beliefs (identified regulation); and enjoyment, curiosity, exploring new ideas (intrinsic motivation).}
\end{figure*}

Our analysis reveals a blend of \emph{extrinsic} and \emph{intrinsic} motivators positioned along a continuum of relative autonomy (see Figure~\ref{fig:motivations}). Early entry is often encouraged by extrinsic factors (e.\,g., sponsorship income, visibility, access to projects). Sustained engagement is driven by more self-determined forms of motivation. Interpreted through SDT, KOLs describe trajectories in which external incentives are gradually internalised into self-endorsed values and identities organised around autonomy, competence, and relatedness. These qualitatively distinct forms of motivation are discussed below.

\noindent\textbf{Recalibrating incentives over time.} Consistent with SDT's account of internalisation, many KOLs recount a shift from externally regulated engagement (e.\,g., taking any sponsorship that pays) towards more self-regulated practices in which they reject misaligned opportunities in order to protect their autonomy and credibility:
\begin{quote}\itshape
``The more I progress in this space and the more mature it gets, I have less and less of an appetite to endorse crypto projects or coins.''
\textemdash{} KOL10
\end{quote}

\noindent\textbf{Credibility over cash.} Chasing money erodes credibility; autonomy sustains trust. Several participants explicitly describe self-imposed rules, such as declining projects they would not personally invest in, that prioritise long-term relatedness with their communities over short-term revenues.

\begin{quote}\itshape
``Even when a sponsor approaches me, I only cover it if I would recommend it to my community anyway. Otherwise, it undermines everything I've built.'' \textemdash{} KOL07
\end{quote}

\noindent\textbf{Education as a worthwhile mission.} Many participants emphasise an educational mission and community service, describing satisfaction from helping newcomers avoid mistakes and make sense of a complex, high-stakes ecosystem. This combination points to competence (mastery and explanation) and relatedness (care for an imagined community of followers).

\begin{quote}\itshape
``Educating people that they have to invest, otherwise they will be broke and poor, is the biggest motivation behind my work.'' \textemdash{} KOL01
\end{quote}

\noindent\textbf{Ideological commitment and mastery.} Interviewees link opinion leadership to personal values (e.\,g., critiques of incumbent finance, decentralisation ideals) and to curiosity and mastery in the fast-moving domain. These narratives foreground autonomy, as participants emphasise self-endorsed commitments to being ``at the forefront'' of technological change rather than simply chasing short-term gains.
Nine participants explicitly highlighted the dynamic nature of the crypto market as a field for impact:
\begin{quote}\itshape
``My motivation was to be at the forefront of this new technological revolution and have a chance to shape it.''
\textemdash{} KOL11
\end{quote}

\noindent\textbf{Key takeaway (RQ1).} Initial extrinsic incentives give way to more internalised motivations. As expertise and recognition grow, KOLs increasingly foreground autonomy (editorial independence and self-imposed standards), competence (continuous learning and translation work), and relatedness (educational stewardship of their communities) as reasons to remain active in high-risk crypto markets, even when this constrains short-term financial opportunities.

\subsection{RQ2: Defining Characteristics of Crypto KOLs}\label{sec:rq2}

We outline characteristics as themes aligned with SDT needs (cf. \Cref{tab:thematic-tree}), showing how autonomy, competence, and relatedness are enacted in high-risk crypto ecosystems and under algorithmic visibility constraints.

\noindent\textbf{Autonomy.}
Autonomy is often expressed in an individual's decision to take a deep dive into crypto- and blockchain-related topics and activities through a self-directed learning path. This process typically involves intensive reading, hands-on experimentation and asset trading, active browsing, and engaging in conversations with others in the community. Most participants described themselves as digital natives and reported technical, financial, or consulting backgrounds, which facilitated both their onboarding and continued interest in this field. Over time, few KOLs have introduced premium content and community memberships as strategies to achieve greater (financial) autonomy and to transition from platform-controlled monetisation to community-driven value creation and exchange.

Once excellence in subject matter and a follower base are secured, KOLs place a high value on the freedom to articulate their own perspectives and share their opinions with a broader audience:
\begin{quote}\itshape
``[As a KOL], you have the influence of letting people know what you think. I can curate an official crypto narrative and put it in the way I feel like it's the right way.'' \textemdash{} KOL01
\end{quote}

Autonomy is also expressed through self-imposed policies and guiding principles that many of the interviewed participants developed to govern their content creation and sponsorship decisions. In particular, crypto KOLs tend to be highly-selective, prioritising value-adding, meaningful, and solution-oriented media publications. By contrast, non-serious or controversial posts are perceived as harmful to authentic self-expression:
\begin{quote}\itshape
``I don't like putting out content to entertain.'' \textemdash{} KOL10
\end{quote}

In rare cases, the autonomy of crypto KOLs extends beyond individual self-expression towards actively shaping the broader institutional environment. By positioning themselves as intermediaries between regulators and the crypto community, KOLs demonstrate both autonomy and competence, thereby aligning their self-endorsed values with influence at the systemic level:
\begin{quote}\itshape
``When it comes to crypto regulation, we lead regulators down a path that is sustainable, because they are often lacking technological know-how. We exert some influence on the ecosystem---not so much on the masses, but indirectly through contributing to the formation of regulation.'' \textemdash{} KOL11
\end{quote}


\noindent\textbf{Competence.}
Crypto KOLs tend to maintain an enduring commitment to developing highly-specialized expertise, often by concentrating on a specific domain or a narrower subset of topics (e.\,g., financial advisory, the dynamics between macroeconomics and cryptocurrencies, crypto-asset taxation, prediction or regional markets). Their competence is primarily demonstrated through the ability to simplify and communicate complex issues in an accessible, timely, and transparent manner:

\begin{quote}\itshape
``If you want to be trusted, you have to be able to articulate quite complicated things in a relatively simple fashion, because blockchain [technology] is a multidisciplinary, quite complicated matter. Nobody can profess you to know it all, but you have to be able to communicate well about all the little bits and pieces that you slice out per your preference. '' 
 \textemdash{} KOL08
\end{quote}

Some KOLs, in particular those offering  advisory services, further perceive themselves as curators and interpreters of financial information, synthesizing multiple sources (e.\,g., on-chain and market data, technical documentation, analytical reports, research publications) and data points to provide clarity and actionable guidance for their followers. These skills distinguish crypto KOLs from other influencers, as their authority is grounded not in popularity, entertainment or promotion value, but in demonstrable expert knowledge, analytical rigour, and the capacity to translate highly technical content into practical insights in a volatile, high-risk environment. In SDT terms, competence is experienced not only as private mastery but as the ability to help others navigate complex infrastructures without resorting to hype.

Beyond knowledge creation and dissemination, crypto influencers invest considerable effort into proactive management of their reputation and personal image. As KOL08 described, maintaining their \emph{social reputation} involves  transparent communication, careful selection of topics and partnerships, and adherence to self-imposed ethical standards. Ultimately, a KOL's reputation serves as a signal of expertise, reinforcing the credibility and trustworthiness of the online content they share.

In terms of content creation strategies, crypto KOLs note that blending educational  materials and investment ideas with emotion-triggering narratives (e.\,g., personal stories, analogies, or provocative framings) elicit the strongest responses from their audiences. This balancing act between education and engagement also mirrors the competitive dynamics of the broader influencer market, in which celebrities and influential figures compete for visibility and audience attention~\cite{franck2019economy}:

\begin{quote}\itshape
``There is a fight for attention, but it's not hostile; people focus on a topic and aim to attract attention through quality content rather than by undermining others.'' 
 \textemdash{} KOL11
\end{quote}

\noindent\textbf{Relatedness.} 
Given the positive effect of interpersonal connections on human well-being~\cite{Reis2000DailyWellBeing}, crypto KOLs actively pursue a sense of belonging and social recognition through their online and offline interactions with like-minded followers. Some invest considerable effort into community building and mutual support, striving to cultivate online environments that foster connectedness, shared purpose, and collective enthusiasm. Others adopt a more conservative approach, prioritising in-person connections and conference participations while using social media primarily as a post hoc amplifier of their offline activities. 

Building on these modes of interaction, our participants also highlighted the importance of strategically navigating through various social media platforms to maintain their influence and visibility. Crypto KOLs differentiate their use of platforms, leveraging Instagram as a medium for outreach and audience acquisition, while employing YouTube for extended content formats with in-depth know-how:
\begin{quote}\itshape
``On YouTube, the community is usually smaller but the content is longer, more binding, and has a strong opinion-shaping effect. By contrast, on X (formerly Twitter) or Telegram the exchanges are made up of short, regular inputs signals that quickly fade away. If people actually engage with YouTube, you have a much stronger influence on them than through these fleeting messages.''  \textemdash{} KOL11
\end{quote}
Despite its declining popularity, X (formerly Twitter) continues to serve as an important channel for many of the interviewed participants, primarily due to its established role in the ecosystem, the concentration of crypto-interested communities, and networking opportunities. At the same time, KOLs increasingly adopt Telegram and Discord, which are regarded as platforms offering more direct community interaction, private modes of engagement, and an access to niche audiences. Many interviewees described TikTok as the least favourable platform, largely due to its popularity among young audiences and the frequent promotion of meme coins and speculative projects. Avoiding such environments was portrayed as a deliberate choice, allowing the interviewed KOLs to protect their professional credibility. 

While Telegram, Discord, and X are generally associated with instant and high-volume exchanges, some interviewees emphasised that LinkedIn fulfils a different role within their communication strategies. In particular, LinkedIn distinguishes itself through its perceived professionalism and the prevalence of authentic, non-fake accounts. For many KOLs, maintaining a presence on LinkedIn complements their credibility-building strategies, as it aligns with their aspiration to be recognised as trustworthy experts rather than mere entertainers. Furthermore, LinkedIn enables crypto KOLs to reach a different audience (e.\,g., companies, entrepreneurs, and developers) and engage in more high-quality discussions compared to the community-driven spaces such as Telegram or Discord. At the same time, several participants noted that LinkedIn's dynamic algorithms require continuous adaptation of content and framing to sustain impressions and ensure that their messages reach intended audiences, highlighting how platform-level engagement metrics shape day-to-day communication choices.

Finally, beyond the platform diversification, KOLs also employ specific communicative practices to reinforce their credibility and strengthen audience ties. Beyond consistent publishing crypto-related materials,  participants described a range of dialogic interactions as crucial for sustaining trust. Common practices include replying to comments, engaging in respectful debates, hosting live Q\&A sessions, and organising webinars and podcasts. Such interactive formats invite feedback and collective sensemaking, thereby strengthening bonds with online communities.


\noindent\textbf{Key takeaway (RQ2).} Authoritative KOLs align autonomy (independence, self-imposed rules), competence (specialised expertise, translation, curation, reputation), and relatedness (community stewardship, dialogic practices, platform navigation) while selectively engaging with platform algorithms and attention dynamics. They differentiate between channels, avoiding spaces that normalise meme-coin speculation and tailoring formats to sustain influence and credibility under shifting visibility metrics.

\subsection{RQ3: Ethics, Social Responsibility, and Community}\label{sec:rq3}

Our participants articulated a nuanced sense of ethical duty and community accountability in their roles as blockchain influencers. They acknowledged that their communication can meaningfully sway token prices and investment sentiment and therefore requires careful self-regulation and transparent disclosure in thin, speculative markets where low-cap tokens can move sharply on limited liquidity. Read through SDT, these accounts show how autonomy (self-imposed rules), competence (bounded claims and careful translation), and relatedness (care for followers' outcomes) underpin KOLs' ethical self-understandings.

\noindent\textbf{Self-restraint and responsible promotion.} Several KOLs recognise that their ability to move markets obliges restraint.  KOL01 explained that their follower count gives them power to influence low-cap tokens and thus obliges them to avoid fraudulent practices:  
\begin{quote}\itshape
``It's definitely my responsibility to not recklessly inside trade.  I noticed I can move an asset that has a market cap of 200{,}000~(USD)—I can move it up $10\text{--}20\%$ easily.''  \textemdash{} KOL01
\end{quote}
KOL03 similarly reflected on the past episode in which they promoted an adult-content project that turned into a publicity disaster.  They now refuse requests to shill tokens\footnote{In the context of cryptocurrency discourse, \emph{to shill} denotes the act of aggressively or uncritically promoting a digital asset, often with undisclosed financial incentives or conflicts of interest. The term carries a negative connotation, implying manipulative or deceptive endorsement rather than impartial evaluation.}, noting that short-term gains would damage their reputation:  
\begin{quote}\itshape
``I have still sometimes requests \ldots would we give you some token?  Could you please push a bit?  I don't, because this actually harms my image.''  \textemdash{} KOL03
\end{quote}
KOL9 and KOL11 avoid direct market commentary altogether.  The former stated that their organisation only posts after results and ``\emph{don't shill tokens, we educate users about the projects},'' while the latter emphasised that as a regulatory expert they refrain from opining on whether particular coins are scams, positioning themself as a neutral actor rather than a promoter. Taken together, these practices exemplify self-regulation: KOLs voluntarily constrain their own promotional options to protect both their reputations and their communities' financial well-being.

\noindent\textbf{Transparency, disclosure, and honesty.} Most interviewees describe strict disclosure norms.  KOL10 recounted that during the NFT boom they labelled every sponsored thread as paid content and that this transparency protected them from backlash when projects underperformed: 
\begin{quote}\itshape
``I would disclose that, hey, this is paid \ldots and it would have market impact \ldots I never got any blame, because if you are upfront and honest \ldots had I not disclosed, I probably would have gotten blamed.'' \textemdash{} KOL10
\end{quote}
They added that they have ``less and less of an appetite'' to recommend specific assets and now stick to high-level portfolio allocations, holding themself to ``a much higher standard.''  KOL2 characterised themself as ``too honest'' and noted that they always warn followers about risks, even though ``people always blame you as an influencer'' regardless of the outcome.  KOL12, representing a venture fund, stated that they will always disclose when a post concerns a portfolio company: 
\begin{quote}\itshape
``We will always write a disclaimer … you have to disclose that you are an investor \ldots this is part of our policy.''  \textemdash{} KOL12
\end{quote}
These statements underscore a shared conviction that undisclosed conflicts of interest violate community norms, yet they also reveal tensions between transparency and accountability: even when disclosures and risk warnings are explicit, influencers anticipate being blamed if followers incur losses.

\noindent\textbf{Community norms and trust building.} Influencers link ethical practice to long-term relationship building.  KOL08 argued that trust hinges on the ability to translate complex concepts without feeding speculation:  
\begin{quote}\itshape
``If you want to be trusted, you have to be able to articulate quite complicated things in a relatively simple fashion \ldots I explicitly never engage in any sort of commentary around the markets.''  \textemdash{} KOL8
\end{quote}
KOL9 emphasised authenticity through ``\emph{meeting people in real life}'' and being ``super doxxed'' in order to show that they are ``really honest'' about what they do.  KOL13 recalled that leaving a traditional finance job was partly motivated by the sector's openness; they highlighted that in the early days ``\emph{everyone was helping each other}'' and that they have ``not seen that good of a knowledge transfer in Web2.''  Such accounts portray a community that values mutual support and education over hype and illustrates how relatedness is sustained through mutual visibility and knowledge sharing.

\noindent\textbf{Managing controversies and reflective practice.} Participants describe strategies for navigating controversies and learning from mistakes.  When asked about a controversial presale, KOL10 explained that they chose not to respond publicly because defending themself would only inflame anger.  KOL3's reflection on their failed promotion taught them to decline paid advertisements and to prioritise social impact over money.  They summarised their credo succinctly:  
\begin{quote}\itshape
``Trust is the most important thing.  How can I promote the trust technology when I cannot be trusted \ldots I don't do any paid ads anymore \ldots it harms my image.''  \textemdash{} KOL3
\end{quote}
KOL13, who regularly publishes long-form research threads, observed that even accurate analyses can be misconstrued as buy signals.  They now avoid validating their own calls and resist the role of a prognosticator:  
\begin{quote}\itshape
``I don't think I've ever revalidated my own thesis $\ldots$ I don't want to end up being a KOL on Twitter at all.  I want to have like 300–400 followers $\ldots$ because then I'll be doing back and forth with [researchers], and people would realize \ldots that's the contradiction we live in---people don't want me to make sense, they want me to give advices.''  \textemdash{} KOL13
\end{quote}
They added that even prominent figures ``\emph{not even Vitalik Buterin\footnote{Vitalik Buterin is a Russian–Canadian programmer best known as the co-founder of Ethereum, one of the most widely used blockchain platforms.} \ldots not even the best traders}'' can reliably pick winning tokens; thus ``\emph{everyone gets to start from the same line},'' reinforcing the need for humility. Such reflections illustrate ongoing internalisation of ethical standards, as KOLs adapt their practices to align more closely with their self-perception of being trustworthy educators rather than forecasters.

\noindent\textbf{Key takeaway (RQ3).} KOLs position themselves as custodians of norms in high-risk, attention-driven markets, enacting credibility through four self-determined practices: (i)~\emph{self-regulation and voluntary constraint}, whereby KOLs decline misaligned sponsorships and impose personal rules on promotion; (ii)~\emph{bounded epistemic competence}, acknowledging limits and avoiding prognostication; (iii)~\emph{accountability}, cultivating long-term trust through transparent disclosure; and (iv)~\emph{reflexive self-correction}, learning from failures and continuously reassessing practices. These practices reframe credibility as an ongoing, ethically enacted performance rather than possessing a set of static credentials.


\section{Discussion}\label{sec:discussion}

Our study sheds light on the diverse motivations, beliefs, and mechanisms of influence that characterise crypto KOLs operating within the Web3 ecosystem. Beyond describing  individual practices, these findings carry broader implications for theories of opinion leadership as well as for the design of socio-technical systems that mediate visibility, trust, and regulation in digital finance. To situate our contributions, we structure the discussion along two perspectives: (1) theoretical implications, where we extend understandings of perceived credibility and its management under conditions of uncertainty and volatility, and (2) practical implications, where we outline design recommendations for social media platforms and supervisory authorities seeking to foster trustworthy, transparent, and community-oriented practices. We further reflect on the application of LLMs in qualitative coding and discuss limitations as well as directions for future research.

\subsection{Theoretical Implications}

One important theoretical contribution of our study is to show that the professional trajectories and day-to-day practices of crypto KOLs can be effectively interpreted through the lens of SDT. This framework has enabled us to examine how KOLs sustain long-term motivation, influence, and legitimacy within decentralised, high-stakes ecosystems characterised by shared financial exposure with followers, irreversible transactions, and thin liquidity. Thus, our findings provide evidence consistent with SDT's applicability by illustrating how the fundamental needs for autonomy, competence, and relatedness manifest in financially risky, technologically mediated, and algorithmically curated environments that differ from the domains where SDT has traditionally been applied~\cite{olafsen2025crafting,sharifi2024basic,beierlein2025creating}. Autonomy is exercised through selective sponsorships, transparent financial disclosures, explicit refusal of misaligned deals, and the rejection of promotional opportunities that could undermine credibility. Competence is demonstrated by simplifying complex knowledge, bounding claims in the face of uncertainty, producing educational content, and cultivating reputations as trusted sources of expertise. Relatedness emerges through dialogic interactions, mutual knowledge sharing, community-building practices, and maintaining social bonds across online and offline contexts.

Second, our findings reconceptualise \emph{credibility} in the Web3 influencer space. Prior research has treated finfluencer credibility as a function of static credentials, namely formal education, professional background, and financial certifications~\cite{NurhandayaniWhoDeserves2025}, reflecting broader source credibility traditions that emphasise expertise and trustworthiness as sender attributes~\cite{Metzger2013CredibilityReview,Wathen2002Credibility,Sundar2008MAIN,Flanagin2007UGC,fogg2001credibility}. Our findings challenge this view. In high-risk crypto environments characterised by widespread misinformation~\cite{Cernera2023rugpulls}, pervasive scepticism, and financial and security risks~\cite{Si2024CHI}, credibility emerges not from credentials but from \emph{self-determined, ethically enacted practices} rooted in SDT needs. We identify four dimensions of this practice-based credibility that function as community-recognised markers of trustworthiness:

\begin{enumerate}[leftmargin=*, nosep]
  \item \textbf{Self-regulation and voluntary constraint:} KOLs decline misaligned sponsorships, impose personal rules on what they will promote, and reject short-term monetisation that could erode trust (Section~\ref{sec:rq3}).
  \item \textbf{Bounded epistemic competence:} Rather than claiming omniscience, KOLs acknowledge the limits of their expertise, avoid prognostication, and emphasise that ``not even the best traders'' can reliably pick winners.
  \item \textbf{Accountability through transparency:} KOLs cultivate long-term trust through consistent disclosure of sponsorships, portfolio holdings, and conflicts of interest, accepting that transparency invites scrutiny.
  \item \textbf{Reflexive self-correction:} KOLs learn from past failures, publicly reassess their practices, and iteratively refine their norms, as illustrated by participants who abandoned paid advertisements after reputational setbacks.
\end{enumerate}

\noindent These dimensions re-conceptualize credibility as a \emph{socio-technical, self-regulated performance} rather than a static credential. Credibility is enacted through ongoing behavioural signals, reinforced by community feedback, and embedded in platform affordances (e.\,g., disclosure badges, accuracy indicators). This perspective extends source credibility theory into high-risk, decentralised contexts and offers a framework for understanding trust in environments where formal qualifications are absent, unverifiable, or irrelevant.

\subsection{Design Recommendations}

Our findings yield actionable insights for platforms, creators, and regulators seeking to foster credible communication and responsible innovation within the evolving Web3 ecosystem. While SDT illuminates why crypto KOLs pursue autonomy, competence, and relatedness, the implications here highlight how such motivations can be translated into governance, practice, and policy design.

\paragraph{Platforms: Designing for credibility and accountability.}
Platforms occupy a pivotal role in shaping socio-technical conditions for credibility. Moving beyond popularity and engagement counts, they should implement multidimensional \emph{credibility metrics} that reward transparent practices, consistent disclosures, and educational rather than sensationalist content. Practical affordances include badges for verified sponsorships or investment statements, indicators that track predictive accuracy over time, and feedback systems enabling users to flag misleading or undisclosed promotions. Equally critical is addressing the fragmentation of identity across channels: interoperable “trust profiles” could consolidate disclosure histories, accuracy scores, and community feedback into a portable reputation system. Coupled with analytical tooling—such as automated detection of undisclosed promotions, sentiment analysis, and cross-platform influence mapping—these infrastructures would make credibility more transparent and comparable, assisting users in distinguishing between hype and substantive guidance while providing structural incentives for ethical behaviour.

\paragraph{Creators and communities: Professional norms and self-regulation.}
Crypto KOLs already exhibit forms of self-regulation, including refusing misaligned promotions, labelling sponsored content, and avoiding financial fraud or similar market manipulation schemes. Formalising these practices into a community-endorsed \emph{finfluencer charter} would provide a normative benchmark for responsible engagement. Such a charter could specify disclosure formats, risk warnings, and principles for mitigating potential market impact. Beyond codification, community-governed mechanisms—peer-review panels, deliberation forums, or recognition systems led by experienced KOLs—can adjudicate contested cases, surface ethical dilemmas, and publicly acknowledge exemplary conduct. These practices would not only guide new entrants but also embed a culture of accountability that scales with the growth of Web3 communities, complementing platform infrastructures and regulatory frameworks.

\paragraph{Policy and regulation: Safeguarding users in volatile markets.}
In light of the transnational and decentralised character of crypto assets, formal oversight remains challenging, yet the potential for retail harm necessitates robust intervention. Regulators could extend existing financial promotion and consumer protection rules to encompass crypto influencers while adapting them to blockchain’s distinctive features. Measures might include standardised disclosure statements for compensated content, registration requirements for actors surpassing defined thresholds of influence, and coordinated enforcement through entities such as the International Organization of Securities Commissions (IOSCO) or regional consortia; guidance from the US Federal Trade Commission (FTC) on endorsements and testimonials, the US Securities and Exchange Commission (SEC) on crypto asset securities, the UK Financial Conduct Authority (FCA) and Advertising Standards Authority/CAP on finfluencer promotions, and the EU’s European Securities and Markets Authority (ESMA) offer relevant precedents and enforcement levers ~\cite{FTC2023EndorsementGuides, SECInvestorCrypto, FCA2023Finfluencer, ASAInfluencer2023, IOSCO2023Crypto, ESMA2023MiCAR}. Platforms can assist by identifying persistent offenders and sharing intelligence with regulators. Alongside enforcement, public agencies should invest in educational initiatives that build crypto literacy among retail investors, reducing susceptibility to misleading guidance. Taken together, harmonised regulation and proactive education would help mitigate risks while supporting responsible innovation.

Collectively, these implications extend beyond individual influencer strategies to encompass the design of accountability infrastructures, cross-platform reputation systems, community norms, and regulatory frameworks. By recognising credibility as a socio-technical outcome that demands coordinated action from platforms, creators, communities, and regulators, the Web3 ecosystem can progress toward more trustworthy and inclusive communication.

\subsection{Reflection\textit{} on LLMs}
In terms of methodology, this study is among the first to empirically demonstrate how LLMs can support qualitative coding in HCI research without displacing human oversight and interpretation. Our hybrid approach aimed to preserve interpretive rigour while reducing the cognitive and temporal load associated with labour-intensive thematic analysis. In our workflow, the LLM operated strictly upstream, suggesting additional candidate codes and SDT mappings on anonymised transcript segments for human review, rather than labelling data directly or participating in reliability calculations. Our experience shows that LLMs can assist human coders in surfacing initial themes, identifying semantic overlaps, and broadening the space of codes under consideration. Nevertheless, they remain prone to generating overly general, repetitive, or normatively loaded suggestions, justifying the need for a dual approach to coding in which humans filter, refine, and ultimately decide which labels enter the codebook. Their use also raises important questions about transparency, reproducibility, ethics, and data privacy~\cite{Schroeder2025LLMs}; we mitigated some of these concerns through strict anonymisation and human-in-the-loop validation, but model biases and versioning remain limitations. For the HCI community, these findings underscore the need for research tools and workflows that integrate LLMs responsibly, ensuring they augment rather than compromise the interpretive depth of qualitative inquiry.

\subsection{Limitations}

This study exhibits several limitations that constrain the generalisability of its findings. First, the sample comprised thirteen self-selected crypto KOLs and leaned Western, with many participants operating within European and U.S. contexts; nevertheless, the cohort also included Asian representation. Additionally, all confirmed participants were male despite active efforts to recruit female KOLs, reflecting the male-dominated composition of the crypto influencer space. Although purposive recruitment afforded heterogeneity in roles and experiences, the modest sample size, geographic skew, and lack of gender diversity restrict the extent to which the observations may be extrapolated to the wider population of crypto KOLs. Second, the reliance on semi‑structured interviews means that data are self‑reported and subject to recall and social desirability biases; participants' narratives were not cross‑validated against behavioural indicators such as on‑chain transaction records or follower engagement statistics. Third, the analysis focused exclusively on influencers' perspectives; the views of followers, regulators, and other ecosystem stakeholders were not solicited, limiting the ability to triangulate perceptions of credibility and influence. Fourth, while LLM‑assisted coding expedited early stages of thematic analysis by suggesting additional candidate labels, it may still have introduced biases inherent in large language models; despite our filtering and human verification steps, model outputs remain shaped by opaque training data and cannot substitute for human interpretive judgement. Fifth, recruitment likely exhibits survivorship and self‑selection bias: KOLs with stronger reputations, clearer disclosure norms, or more reflective practices may have been more willing to participate, potentially underrepresenting highly promotional or opaque actors. Finally, the inherent volatility of crypto markets and the rapid evolution of Web3 platforms imply that the practices and norms documented here may be transient, underscoring the need for longitudinal replication.

\subsection{Future Research Directions}

Future investigations should address these limitations and advance the field in several directions.  Comparative and cross‑cultural studies drawing on larger and more diverse samples would help to assess the universality of the present findings and reveal how regulatory and cultural contexts shape influencer behaviour.  Longitudinal mixed‑methods designs that integrate interviews with trace data—such as social media analytics, on‑chain transactions, and algorithmic amplification metrics—could illuminate how credibility develops over time and how KOLs' recommendations correlate with market dynamics.  Incorporating follower perspectives through surveys and ethnographic observation would elucidate parasocial interactions and audience responses to disclosure practices.  Methodological innovation is warranted to develop and validate computational measures of credibility, transparency, and influence that enable systematic cross‑platform assessment; such tools could draw on machine learning and natural language processing to detect undisclosed promotions and sentiment.  Examining the implications of emerging technologies—such as generative AI and decentralised identity infrastructures—for content production, reputation management, and compliance represents another important avenue.  Finally, comparative analyses across blockchain and other influencer domains could help distinguish domain‑specific behaviours from broader trends in the creator economy.


\section{Conclusion}\label{sec:conclusion}

This study has examined how crypto key opinion leaders understand their motivations, credibility, and responsibilities. We conceptualise KOLs as hybrid figures: educators, informal financial advisors, entrepreneurs, and community stewards, whose credibility emerges from motivational orientations, cross-platform and on-chain practices, and collectively enforced norms. Drawing on self-determination theory and interviews with 13 KOLs, we showed how autonomy, competence, and relatedness needs are negotiated alongside monetisation pressures, token holdings, and shared financial exposure with followers. Our hybrid human--LLM workflow for thematic analysis demonstrates the potential of large language models to broaden candidate theme coverage while underscoring the necessity of human oversight to mitigate hallucinations and hidden biases.

\textbf{Regulatory implications.}
These insights carry practical implications at a moment when influencer regulation around financial content is tightening globally, with jurisdictions such as China, the European Union, and Singapore introducing stricter oversight of crypto-related promotion \cite{ESMA2023MiCAR,MAS2024crypto}. Our findings suggest that effective governance will need to account for the blurred boundaries between education, promotion, and speculation that define crypto KOL work. Rather than treating KOLs simply as advertisers or financial advisors, regulators may benefit from recognising the hybrid nature of their roles and the community dynamics that sustain their influence. Designing mechanisms that support transparent incentives, encourage responsible signalling, and reduce asymmetries of financial exposure (such as the disclosure badges, accuracy indicators, and interoperable trust profiles discussed in Section~\ref{sec:discussion}) may strengthen both creator credibility and user protection as crypto markets continue to evolve.
\begin{acks}
    This research was funded by the FFG project Finfluencer (grant agreement number 924721). This paper was accepted at CHI 2026.
\end{acks}

\bibliographystyle{ACM-Reference-Format}
\bibliography{references}

\appendix \label{app1}
\section{Thematic Framework}

As an overview of our synthesis, \Cref{tab:thematic-tree} presents the thematic tree linking overarching themes and subthemes to the research questions, alongside concise definitions and representative quotations.

\begin{table*}[p]
    \centering
    \footnotesize
    \setlength{\tabcolsep}{3pt}
    \renewcommand{\arraystretch}{1.25}
    \caption{Thematic tree: overarching themes and subthemes mapped to the research questions (RQ).}
    \label{tab:thematic-tree}
    \begin{tabular}{@{} p{2.2cm} p{3cm} p{4.8cm} p{4.8cm} c @{}}
    \toprule
    \textbf{Overarching Theme} & \textbf{Subtheme} & \textbf{Definition ($\leq$20w)} & \textbf{Representative Quote} & \textbf{RQ(s)} \\
    \midrule
    Autonomy & Editorial independence & Choosing topics/sponsors to preserve credibility and control. & ``I only cover it if I'd recommend it to my community.'' (KOL07) & RQ2 \\
    Autonomy & Self-directed learning & Deep dive via reading, experimenting, trading, conversations. & ``I can curate a narrative and put it the right way.'' (KOL01) & RQ2 \\
    Autonomy & System-shaping advocacy & Influencing regulators/ecosystem in line with values. & ``We lead regulators down a sustainable path.'' (KOL11) & RQ2 \\
    \midrule
    Competence & Translating complexity & Make technical topics clear without oversimplifying. & ``Articulate complicated things in a simple fashion.'' (KOL08) & RQ2 \\
    Competence & Evidence curation & Synthesize on-chain, docs, reports for guidance. & ``People focus on quality content to attract attention.'' (KOL11) & RQ2 \\
    Competence & Reputation management & Transparent standards signal expertise and trust. & ``Social reputation'' via careful topics/partnerships. (KOL08) & RQ2 \\
    \midrule
    Relatedness & Community stewardship & Education and long-term ties over hype. & ``Everyone was helping each other.'' (KOL13) & RQ3 \\
    Relatedness & Dialogic practices & Comments, debates, Q\&A, webinars, podcasts. & ``Meeting people in real life\ldots be `super doxxed'.'' (KOL09) & RQ3 \\
    Relatedness & Platform navigation & Match platforms to audiences and norms. & ``LinkedIn felt more authentic, professional.'' (Synth. from interviews) & RQ2,RQ3 \\
    \midrule
    Motivation & Intrinsic—education/altruism & Desire to inform, raise literacy, serve community. & ``Educating people\ldots is the biggest motivation.'' (KOL01) & RQ1 \\
    Motivation & Intrinsic—curiosity/mastery & Enjoyment, curiosity, shaping a new field. & ``Be at the forefront of this revolution.'' (KOL11) & RQ1 \\
    Motivation & Extrinsic—sponsorship/visibility & Early income/visibility; wanes as values internalize. & ``Less appetite to endorse coins over time.'' (KOL10) & RQ1 \\
    \midrule
    Credibility Practices$^{\dagger}$ & Self-regulation \& voluntary constraint & Decline misaligned sponsorships; impose personal rules on promotion. & ``I don't\ldots because it harms my image.'' (KOL03) & RQ3 \\
    Credibility Practices$^{\dagger}$ & Bounded epistemic competence & Acknowledge limits; avoid prognostication and overclaiming. & ``Not even Vitalik\ldots not even the best traders.'' (KOL13) & RQ3 \\
    Credibility Practices$^{\dagger}$ & Accountability \& transparency & Clear sponsorship/investment disclaimers; accept scrutiny. & ``I would disclose that\ldots this is paid.'' (KOL10) & RQ3 \\
    Credibility Practices$^{\dagger}$ & Reflexive self-correction & Learn from controversies; continuously reassess practices. & ``Trust is the most important thing\ldots no paid ads.'' (KOL03) & RQ3 \\
    \bottomrule
    \addlinespace[2pt]
    \multicolumn{5}{@{}l}{\footnotesize$^{\dagger}$\,These four dimensions constitute self-determined credibility practices that function as community-recognised markers of trustworthiness.}
    \end{tabular}
    \end{table*}

\section{Literature Synthesis}

Table~\ref{tab:lit-synthesis} provides a structured overview of the key literatures informing our study, their central constructs and findings, their relevance to understanding crypto KOLs, and the gaps that remain.

\begin{table*}[p]
\centering
\caption{Synthesis of prior research streams. 
}
\label{tab:lit-synthesis}
\footnotesize
\renewcommand{\arraystretch}{1.5}
\setlength{\tabcolsep}{3pt}
\begin{tabular}{@{} p{3cm} p{3cm} p{3.8cm} p{3.6cm} p{3.5cm} @{}}
\toprule
\textbf{Research Stream} & \textbf{Key Constructs} & \textbf{Central Findings} & \textbf{Relevance to Crypto KOLs} & \textbf{Gap Addressed} \\
\midrule
Influencer culture \& parasociality \newline {\cite{MarwickStatusUpdate2013,AbidinVisibility2016,Lou2019,Labrecque2014,Reinikainen2020,Khamis2017,Enke2019,BaymPlaying2018}}
  & Visibility labour; relational labour; authenticity; parasocial interaction
  & Credibility cultivated through curated authenticity and affective bonds; trust based on perceived closeness
  & KOLs engage in similar visibility and relational labour; credibility central to influence
  & Does not address financial stakes or speculative contexts \\[6pt]
Finfluencers \& social-media finance \newline {\cite{GuanFinfluencer2023,HayesFinfluence2024,HaaseFinfluencersSentiment2025,MeyerHighOnBitcoin2023,StefanouFinfluencersReg2022}}
  & Finfluencer; emotional contagion; sentiment transfer; financialisation
  & Finfluencers shape crowd sentiment; emotional contagion amplifies herd dynamics
  & KOLs are crypto-specific finfluencers with intensified risks
  & Focuses on audiences and markets, not on finfluencers as workers \\[6pt]
Creator labour \& financialisation \newline {\cite{DuffyAspirationalWork2017,DuffyNestedPrecarities2021,DuffyPlatformPractices2019,CotterVisibilityGame2019,AlacovskaSpeculativeLabour2024,AbidinInternetCelebrity2018}}
  & Aspirational labour; nested precarities; speculative labour
  & Creators face layered precarities; financialisation entangles creative work with volatile markets
  & KOLs experience double exposure to algorithmic and financial volatility
  & Does not examine crypto-specific contexts \\[6pt]
Crypto users \& communities \newline {\cite{MoserBitcoinUsability2013,Krombholz2016,Mai2020,Abramova2021,KhairuddinBitcoinMotivations2016,SasKhairuddin2017,BappyCentralizedTrust2025}}
  & Mental models; usability; centralised trust; user motivations
  & Users hold fragile mental models; paradoxically trust centralised actors in decentralised systems
  & Users ``outsource'' due diligence to KOLs as trust anchors
  & Focuses on users, not intermediaries \\[6pt]
Regulation \& consumer protection \newline {\cite{StefanouFinfluencersReg2022,SECInvestorCrypto,FCA2023Finfluencer,ASAInfluencer2023}}
  & Investor protection; disclosure; regulatory fragmentation
  & Finfluencer promotion evades traditional regimes; crypto regulation is patchy
  & KOLs operate in regulatory grey zones; accountability unclear
  & Does not examine KOL perspectives on regulation \\[6pt]
Scams \& fraud \newline {\cite{OakScamReddit2025,Cernera2023rugpulls}}
  & Rug pulls; scam discourse; community vigilance
  & Scams are pervasive; communities develop grassroots guidance roles
  & KOLs may help or harm users navigating scams
  & Does not examine KOL role or responsibility \\[6pt]
SDT \& online work \newline {\cite{DeciRyan1985,RyanDeci2000,NovMotivatesWikipedians2007,TyackMeklerSDT2020,LiuAnxietyCommunities2021,torhonen2019}}
  & Autonomy; competence; relatedness; intrinsic vs.\ extrinsic motivation
  & SDT needs predict sustained, high-quality engagement; prosocial motives common among creators
  & SDT offers lens on KOL motivations beyond ``greed vs.\ altruism''
  & Has not been applied to crypto influencers \\[6pt]
Market-level influence \newline {\cite{LucchiniGamestopWSB2022,warkulat2024,pan2025,AnteMuskCrypto2023,Merkley2024}}
  & Sentiment; attention; capital flows; price effects
  & Online communities and influencers affect trading volumes and prices
  & KOLs' endorsements can move markets and affect follower wealth
  & Models influencers as signals, not as workers \\
\bottomrule
\end{tabular}
\end{table*}

\section{LLM Prompts and Outputs}

This appendix provides the LLM prompt used for thematic analysis as well as an exemplary excerpt of LLM output.

\begin{figure}[H]
\caption{LLM prompt used for thematic assistance.}\label{lst:ai-prompts}
\Description{A framed code listing showing the system prompt and user prompt used for LLM-assisted thematic analysis, instructing the model to suggest candidate themes mapped to self-determination theory needs (autonomy, competence, relatedness) and return results as a JSON array with theme, description, and SDT category fields.}
\centering
\begin{minipage}{0.82\columnwidth}
\begin{Verbatim}[frame=single,fontsize=\scriptsize]
system_prompt = """You are a qualitative
researcher assisting with thematic analysis
of interview transcripts about crypto-
influencers' motivations. Your task is to:

1. Suggest candidate themes (labels) that
   capture emerging patterns in the text
2. Provide a brief description of each theme
3. Map each theme to one of the three Self-
   Determination Theory (SDT) needs:
   - Autonomy: The need to feel volitional
     and self-directed in one's actions
   - Competence: The need to feel effective
     and capable of achieving desired outcomes
   - Relatedness: The need to feel connected
     to others and experience belongingness

Focus on identifying motivations, behaviours,
and psychological needs expressed by the
participants.

Return your analysis as a JSON array of
objects with these fields:
- "theme": the candidate thematic label
- "description": what the theme means in
  this context
- "sdt_category": one of "Autonomy",
  "Competence", or "Relatedness"

Example format:
[
  {
    "theme": "Financial Independence Seeking",
    "description": "Desire to achieve financial
      freedom through crypto investments and
      avoid traditional employment constraints",
    "sdt_category": "Autonomy"
  }
]
"""

user_prompt = f"""Analyze this interview
transcript excerpt and suggest candidate themes:

Context: Interview transcript from {filename},
chunk {chunk_index + 1}

Text to analyze:
{text_chunk}

Generate 3-7 candidate themes that capture the
key motivations, practices, and needs expressed
in this text segment."""
\end{Verbatim}
\end{minipage}
\end{figure}

\paragraph{LLM transparency note.}
We used the best available GPT-4 model (June 2025) with temperature~$=0.2$ and top\_p~$=1.0$. All excerpts were fully anonymised prior to processing; no raw data or identifiers were transferred externally. Outputs were reviewed and verified by human coders.

For completeness, an example excerpt of the model's JSON output structure is shown in \Cref{lst:ai-output}.

\paragraph{Code curation summary.}\label{app:code-curation}
Table~\ref{tab:code-curation} summarises the code reduction pipeline. The overall rejection rate of LLM-suggested codes (47\%) reflects a conservative curation strategy in which human judgement remained the final arbiter.

\vspace{-2pt}
\begin{table}[H]
\centering
\small
\caption{Code curation pipeline: counts and rejection rates at each stage.}
\label{tab:code-curation}
\setlength{\abovecaptionskip}{4pt}
\setlength{\belowcaptionskip}{2pt}
\begin{tabular}{lcccc}
\toprule
\textbf{Stage} & \makecell{\textbf{Codes}\\\textbf{in}} & \textbf{Retained} & \textbf{Rejected} & \makecell{\textbf{Rej.}\\\textbf{Rate}} \\
\midrule
LLM first pass    & 60  & 32 & 28 & 47\% \\
Human codes added  & +19 & 19 & 0  & ---  \\
Combined pool      & 51  & 36 & 15 & 29\% \\
Final codebook     & 36  & 36 & --- & ---  \\
\bottomrule
\end{tabular}
\end{table}
\vspace{-4pt}

\noindent LLM-suggested codes were rejected for: \textbf{platform-specificity} (4 codes, e.\,g., ``LinkedIn for Professional Use''); \textbf{conceptual conflation} (6 codes, e.\,g., ``Critical Evaluation of Influencers''); \textbf{insufficient grounding} (9 codes, e.\,g., ``Global Connectivity through Gaming''); and \textbf{redundancy with human codes} (9 codes, e.\,g., ``Commitment to Transparency''). Each rejection was logged with a rationale to maintain an auditable curation trail.

\begin{figure}[H]
\caption{Example model output (excerpt).}\label{lst:ai-output}
\Description{A framed code listing showing an example JSON output excerpt from the LLM, containing four candidate themes (Educational Mission, Autonomy in Content Choices, Mastery Through Continuous Learning, Reputation and Recognition) each with a description and mapped SDT category.}
\centering
\begin{minipage}{0.82\columnwidth}
\begin{Verbatim}[frame=single,fontsize=\scriptsize]
[
  {
    "theme": "Educational Mission",
    "description": "Motivation to translate
      complex crypto concepts and help
      newcomers avoid mistakes",
    "sdt_category": "Relatedness"
  },
  {
    "theme": "Autonomy in Content Choices",
    "description": "Emphasis on independence
      from sponsors and freedom to express
      critical views",
    "sdt_category": "Autonomy"
  },
  {
    "theme": "Mastery Through Continuous
      Learning",
    "description": "Drive to keep up with
      fast-moving protocols and improve
      analytical skills",
    "sdt_category": "Competence"
  },
  {
    "theme": "Reputation and Recognition",
    "description": "Seeking status and
      credibility as a knowledgeable KOL
      within the ecosystem",
    "sdt_category": "Competence"
  }
]
\end{Verbatim}
\end{minipage}
\end{figure}

\vfill

\section{Interview Questions}\label{app:interview-questions}

\begin{enumerate}[label=\textbf{\arabic*.}, leftmargin=8mm, nosep]
  \item \textbf{Background and Role in the Ecosystem}
  \begin{enumerate}[label=Q\arabic*., leftmargin=8mm, start=1, nosep]
    \item Can you briefly describe your professional background and how you got involved in the crypto assets industry? \hfill{(RQ2)}
    \item What led you to become an influential figure in the crypto space, and how would you define your role within it? \hfill{(RQ2)}
  \end{enumerate}
  \item \textbf{Motivations and Goals}
  \begin{enumerate}[label=Q\arabic*., leftmargin=8mm, start=3, nosep]
    \item What motivates your active participation in decentralized ledger–related discussions and project endorsements? \hfill{(RQ1)}
    \item Based on your role as a [self-defined ROLE from Q2], what drives you to engage with the crypto space in this particular way? \hfill{(RQ1)}
  \end{enumerate}
  \item \textbf{Influence Mechanisms and Engagement}
  \begin{enumerate}[label=Q\arabic*., leftmargin=8mm, start=5, nosep]
    \item Through what channels do you believe you achieve the most influence (e.\,g., X, YouTube, podcasts, conferences, direct investments), and why? \hfill{(RQ2)}
    \item How do you engage with your audience, and what type of content do you believe generates the most engagement and resonance? \hfill{(RQ2)}
  \end{enumerate}
  \item \textbf{Community and Industry Dynamics}
  \begin{enumerate}[label=Q\arabic*., leftmargin=8mm, start=7, nosep]
    \item How do you interact with other opinion leaders in the decentralized ledger sphere? Would you describe your relationships as collaborative, competitive, or neutral?\hfill
    \item Have you ever engaged in public debates or controversies within the digital ledger space? How do you navigate such interactions? \hfill{(RQ2, RQ3)}
  \end{enumerate}
  \item \textbf{Influence on Investment, Adoption \& Ethics}
  \begin{enumerate}[label=Q\arabic*., leftmargin=8mm, start=9, nosep]
    \item Have you observed cases where your recommendations significantly impacted market trends or investment behaviour? How do you view this responsibility? \hfill{(RQ3)}
    \item How do you ensure transparency and credibility in your public statements, particularly regarding financial incentives and endorsements? \hfill{(RQ1, RQ3)}
  \end{enumerate}
\end{enumerate}



\end{document}